\documentclass[titlepage]{article}

\newcommand{\di}{\ensuremath{\mathrm{d}}}

\title{Factorization in Dual Models and Functional Integration in String Theory.} 

\author{Stanley Mandelstam  \\ Department of Physics,  University of California, Berkeley, \\California 94720, U. S. A. \\and \\Lawrence Berkeley Laboratory, University of California, Berkeley,\\ California 94720, U. S. A.  \thanks{Supported by the U. S. Department of Energy under contract \mbox{DE-AC02-05CH11231}} \vspace{0.4in}\\
Contribution to the volume "The Birth of String Theory",\\ eds. A. Cappelli, E.Castellani, F. Colomo, P DiVecchia \vspace{0.7in}\\
\parbox{5.25in}{This article contains a summary of the author's contributions, one in collaboration with K. Bardak\c{c}i, to dual models and string theory prior to the mid-seventies.  Other workers'contributions,  during and subsequent to this period,  are mentioned in order to relate our work to the general development of the subject.}}

\begin{document} 

\maketitle

\section{Introdution}
In accordance with the scope of the present volume,  this article will mainly be concerned with
work during the relevant period in which the author was involved, with a minimum of technicalities.   Other contributions  will be mentioned briefly  in order to indicate the place of our work in the general development.  Details of these contributions as   will not be given,  as they are adequately covered in other articles in the volume.
We emphasize that the contributions are chosen in order to provide continuity with the contributions in which we were involved,  and not for their importance.   There will be many uncited contributions of greater importance than those cited. The historical development will be emphasized.

As every reader of this volume probably knows, string theory can be traced back to the discovery by Veneziano~\cite{Ven} of a formula for a four-point scattering amplitude (two incoming and two outgoing particles) with narrow resonances and rising Regge trajectories in both channels.  Another such formula,   for what are now known as closed strings,  was proposed by Virasoro~\cite{Vir} and written as an integral by Shapiro~\cite{Sh}. These formulas were then extended  to five-point amplitudes independently by Bardak\c{c}i and Ruegg and by Virasoro~\cite{BRV} and then to $N$-point amlitudes independently by Bardak\c{c}i and Ruegg; Chan and Tsou; Goebel and Sakita; and Koba and Nielsen~\cite{n-point} \cite{KN}. For simplicity most of the treatment in the remainder of this article will be for open strings;  the closed-strings formulas will be very similar.  The most convenient expression for the general open-string $N$-point amplitude is the following,  which we quote here for future reference
\begin{equation}
A=\int {\mathrm d}^{n-3}z|(z_b-z_a)(z_c-z_b)(z_c-z_a)| \prod_{i>j}|z_i-z_j|^{-2p_ip_j}  \label{eq:KN}
\end{equation}
The subscript $i$ refers to the $i$-th particle,  with momentum $p_i$. The variables $z_i$ are ordered cyclically alont the real line.   Three of them,  which we have denoted by $z_a, z_b, z_c$ are arbitrarily chosen and held at arbitrary fixed values; we integrate over the other $n-3$ subject to the condition of cyclic ordering.
By making use of the invariance of the amplitude (\ref{eq:KN}) under a projective transformation of the  variables
\begin{equation}
z'=\frac{az+b}{cz+d},  \label{eq:proj}
\end{equation}
we can easily see that the integral is independent of the choice and values of the constant $z$'s.   We have written the formula for the case were the external particles are tachyons with $\mu ^2$, the square of their mass,  equal to -1 in units of the slope of the Regge trajectories;  the formula for the general case is slightly more complicated.  At this stage of the development there is no reason for this choice of the mass,  but later on we shall see that it is necessary in a consistent model.

The formula for the closed-string amplitude is similar but not identical to (\ref{eq:KN}),  except that the integral is now over the whole complex plane,  and the
variables are not restricted by cyclic ordering.

If the variable $p_i.p_j$ is sufficiently large,  the above integral diverges when $   (z_i-z_j) $ approaches zero.  One can obtain  finite results by starting from an algebraically smaller value of $ p_i.p_j $ and analytically continuing or, equivalently, by integrating by parts and dropping the end-point contribution.  At the time,  such a procedure was justifiable, since one was simply trying to obtain an S-matrix element with the correct analytic properties.  We shall return to this point later.

\section{Factorization}
\setcounter{equation}{0}

In order that the scattering amplitudes should together form a consistent $S$-matrix,  they must satisfy a requirement known as factorization (the narrow-resonance equivalent of unitarity).  The residue at the pole of an amplitude as a function of $s$,  the square of the energy,  should consist af a number of terms,  one for each excited particle, with the square of its mass equal to $s$..  Each term has to be the product ot two factors,  one depending only on the number of  incoming  particles and their momenta, the other on the corresonding variables of the outgoing particles
\begin{figure}

\begin{picture}(468,110)(122,72)

\put(252,124){\circle*{18}}

\put(331,124){\circle*{18}}

\put(173,124){\line(1,0){79}}

\put(196,68){\line(1,1){56}}

\put(196,180){\line(1,-1){56}}

\put(331,124){\line(1,0){79}}
\put(331,124){\line(1,1){56}}

\put(331,124){\line(1,-1){56}}

\thicklines

\put(252,124){\line(1,0){79}}

\end{picture} 

\caption{Factorization of scattering amplitudes at a pole.}

\end{figure}
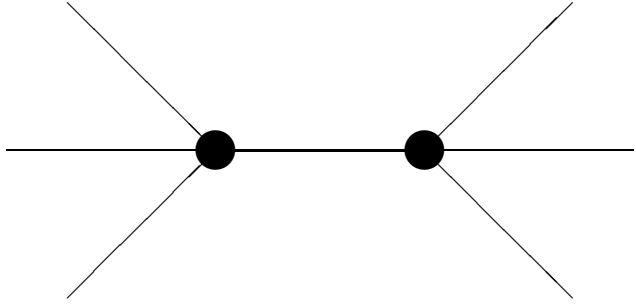
The factorizability of the amplitude (\ref{eq:KN}) was shown independently by Bardak\c{c}i and Mandelstam  and by Fubini and Veneziano ~\cite{BMFV}.  In Figure 1,  we define the variables corresponding to the particles on the left as $ z_1,...z_j $,  and those corresponding to the particles on the right as $ z_{j+1},...z_N $.  We fix the variables $ z_1,z_{j+1}$ and $z_N $ at the values  0, 1, and $\infty $.  We then define new variables
\begin{equation}
x_i=\frac{z_i}{z_j}, \hspace{.18in} 1\leq i\leq j \hspace{.42in} z=z_j, \hspace{.42in} y_k=\frac{z_{j+1}}{z_k}, \hspace{.18in} j+1\leq k\leq N  \label{eq:vars}
\end{equation}
The  formula (\ref{eq:KN}) now becomes
\begin{equation}
A=\int{\mathrm{d}}z{\mathrm{d}}x_2,...{\mathrm{d}}x_{j-1}{\mathrm{d}}y_{j+2},...{\mathrm{d}}y_{N-1}z^{-s-1}I_1
I_2 \prod_{i=1}^{j}\prod_{k=j+1}^{N}(1-zx_iy_k)^{-2p_i.p_k}, \label{eq:KNfact}
\end{equation}
where
\begin{displaymath}
I_1 =\prod_{i=1}^{j} \prod_{k=i+1}^{j} (x_k -x_i)^{-2p_i.p_k}, \hspace{.4in}  I_2
=\prod_{i=j+1}^{N} \prod_{k=i+1}^{N} (y_i -y_k)^{-2p_i.p_k},
\end{displaymath}
\begin{displaymath}
0=x_1\leq x_2...\leq x_j=1, \hspace{.4in}  1=y_{j+1}\geq...\geq y_{N-1}\geq y_N=0,
\end{displaymath}
and $s$ is the square of the energy of the intermediate state in Figure 1.

To calculate the residue $R$ at the pole in (\ref{eq:KNfact}) when $s=n$,  where $n$
is a non-negative integer,  we first have to integrate by parts $n$ times as
described above.
We then find
\begin{equation}
R=-\frac{1}{n!} \int \di x_2 ...  \di x_{j-1} \di y_{j+2} ... \di y_{N-1} I_1I_2
\frac{\partial^n}{\partial z^n} \prod_{i=1}^{j}\prod_{k=j+1}^{N}(1-zx_i y_k
)^{-2p_i.p_k }|_{z=0} \label{eq:KNfact2}
\end{equation}
By expanding the last factor in (\ref{eq:KNfact2}) in a power series and taking the
coefficient of $z^n$,  we find that R can be written as the sum of terms of the form
\begin{equation}
c\int \di x_2 ...  \di x_{j-1} \di y_{j+2} ... \di y_{N-1} I_1I_2\prod_{r}
\sum_{i=1}^{j} \sum_{k=j+1}^{N}(2p_i.p_kx_i^r y_k^r)^{\lambda_r},  \label{eq:KNfact3}
\end{equation}
where
\begin{equation}
\sum r\lambda_r=n  \label{eq:KNfact4}
\end{equation}

Equations (\ref{eq:KNfact3}) and (\ref{eq:KNfact4}) show that the residue is indeed a finite sum, each term corresponding to a state of an assembly of $d$-vector simple harmonic oscillators. The integer $r$ labels the oscillator,  the level spacing of the $r$th oscillator is proportional to $r$,  and the energuy of the ground state is equal ot zero. In the term shown the $r$th oscillator is in level $\lambda_r$.
(More accurately,  we should expand the dot product in (\ref{eq:KNfact3}) and then expand the $\lambda_r$th power of the sum;  each term will correspond to a state where the $r$th oscillator is in level $\lambda_r$,  with the energy divided among the components in the obvious way.)  We have thus shown that the  amplitude (\ref{eq:KN}) does factorize as required,  and the states of the system correspond to the assembly of simple harmonic oscillators described above.

The factorization properties of the $N$-point amplitude were used independently by Fubini, Gordon and Veneziano, by Nambu and by Susslind~\cite{FGV} to construct an operator formalism for the dual model, the operators being the creation and destruction operators for the above assembly of  oscillators.  These authors were able to construct the $N$-point amplitude in terms of such operators;  the factorization properties were then obvious.  In fact, the easiest way to describe dual models was seen to be to start from the operator formalism,  thus reversing the historical order.

\section{Further Developments}
\setcounter{equation}{0}

Two very similar new dual models were proposed by Ramond~\cite{R} (R), initially for
free fermions, and by Neveu and Schwarz~\cite{NS} (NS) for bosons;  the latter
model was slightly reformulated by Neveu,  Schwarz and Thorn.  These models both had
a series of anti-commuting $d$-vector harmonic-oscillator operators ($b$-operators) in
addition to the $a$-operators mentioned at the end of the last section.  The
difference was that the Ramond operators had integral mode numbers,  the zero-mode
operators being interpreted as $\gamma$-matrices,  while the NS operators had
half-integral mode numbers.

It was suggested by Kikkawa, Sakita and Virasoro~\cite{KSV} (KSV) that the 
amplitude  (\ref{eq:KN}) might be the Born term of a perturbation series,  and they proposed a form
for the $n$-loop term.  The simpest one-loop term was calculated independently from
unitarity by
Amati, Bouchiat and Gervais, by Bardak\c{c}i, Halpern and Shapiro and by Kikkawa,
Sakita, Veneziano and Virasoro~\cite{one-loop}.  More general terms were calculated
by by Kaku and Thorn~\cite{KT};  an improvement to their calculation was made by
Gross, Neveu, Scherk and Schwarz (GNSS)~\cite{GNSS}.  The $N$-loop amplitede was
calculated independently by Alessandrini, by Kaku and Yu and by
Lovelace~\cite{n-loop}; their work was further developed by Alessandrini and
Amati~\cite{AA}.  Lovelace and Alessandrini,  in common with all these workers,  based their
calculation on the operator formalism mentioned above, but they pointed out that their
calculation could be understood in terms of amplitudes associated with $n$-hole
Riemann surfaces.  Their work was motivated by ideas suggested by
Nielsen~\cite{string} in connection with the recently proposed  string
interpretation of dual models, and the associated analogue model of Fairlie and
Nielsen~\cite{FN}. All these calculations, however, had ghosts,\footnote{By ghost we
mean any particle with a negative metric.  The Faddeev-Popov ghosts,  of course,
were not involved in dual models or string theory at this stage.} which we shall
treat shortly,  circulating in the loops, and they were therefore incorrect.  We
mention them because of their relation to later work.

GNSS showed that certain one-loop amplitudes possessed cuts not associated with
unitarity.  Such cuts were found independently by Frye and Susskind \cite{FS} in a
calculation based on the analogue model.   Lovelace~\cite{L2},  making a reasonable
conjecture  about the amplitude with ghosts eliminated, showed that these cuts
became poles if $d$=26.  It thus appeared that the model was inconsistent unless
$d$=26, a   feature which caused considerable amusement among sceptics at the time.

Since the products in Eq. (\ref{eq:KNfact3}) are $d$-vector products, or,
equivalently, the $a$-operators in the operator formalism are $d$-vectors, the states
corresponding to the time component will be ghosts, i. e., negative-metric states.
The model will not therefore be physicslly acceptable unless the ghosts can be
eliminated.  In the case where the ground state was a tachyon with $\mu^2=-1$,  Virasoro~\cite{Vir2}
found an infinite series of operators which,  when applied to any particle in the
operator formalism,gave a "spurious" state, i. e., a state which would not appear as
an intermediate state in Figure 1 (The Virasoro operators  satisfiy the algebra of
two-dimensional conformal transformations with a central charge.)  The physical
Hilbert space was therefore the space orthogonal to all the spurious states,  and
the hope was that this Hilbert space would be ghost-free.  Del Giudice, DiVecchia
and Fubini~\cite{DDF} (DDF) found a set of "transverse" positive-definite
simple-harmonic oscillators, with $d-2$ components, which commuted with the
Virasoro operators and which could therefore be used to construct a ghost-free
subspace of the Hilbert space, or possibly the whole Hilbert space.  Finally
Brower~\cite{Bro} showed that the DDF operators, together with a new set of
"longitudinal" operators, could be used to construct the entire physical Hilbert
space,  and that the Hilbert space so constructed  was ghost-free provided $d\leq
26$.  

Brower showed that the situation was partilarly simple if $d=26$, when the
longitudinal operators produced null states which did not give rise to poles in Figure
1,  so that the physical Hilbert space constructed from the DDF operators without the longitudinal operators was
sufficient.  The string models, as presently formulated,  cannot incorporate the
longitudinal modes,  and, in fact, no procedure is known for allowing only the
particles in the physical Hilbert physical space to circulate in loops if there are
transverse and longitudinal modes\footnote{But such a model may possibly be
equivalent to the Polyakov noncritical string}  We thus again obtain the condition
$d$=26 as a condition for consistency. We shall denote $d$=26 as the critical
dimension.  Goddard and Thorn \cite{GT} have given a simpler proof of ghost
elimination if $d$=26.

The ghost-free physical Hilbert space for the NS model can be constructed in a very
similar way~\cite{NSghost}~\cite{GT}.  Here the square of the mass of the ground
state must be equal to minus one-half of the slope of the Regge trajectory, and the
critical dimension is $d$=10.

It was observed independently by Nambu,  Nielsen and Susskind~\cite{string} that the
oscllators of dual models could be considered as the modes of vibration of a string.
Nambu and Goto \cite{Go} took the action for the world sheet of the string moving in
 time to be the area of the sheet.  Goddard, Goldstone, Rebbi and Thorn \cite{GGRT} 
(GGRT) have quantized  the noninteracting dual string.  They first considered a 
classical string and used light-cone coordinates
\begin{equation}
X^+=\frac{1}{\sqrt{2}}(X^0 + X^{d-1}), \hspace{.3in}  X^-=\frac{1}{\sqrt{2}}(X^0 -
X^{d-1}),  \hspace {.3in } X^i \hspace{.15in}  1\leq{i}\leq{d-2}     \label{eq:lqvars}
\end{equation}
They then made a choice of coordinates $\sigma ,\tau$ on the string world-sheet. 
Making such a choice is equivalent to fixing a gauge in a gauge theory.  They showed
that they could make a choice such that $\tau$=$X^+ $,  that the momentum per unit
length $P^+$ is a constant, that $X^-$ is determined in terms of the $X^i$'s, and
that the Nambu-Goto Lagrangian is
\begin{equation}
\mathcal{L}=\frac{1}{4\pi} \left\{ \left( \frac{\partial X^i}{\partial \tau} \right)^2 -  \left( \frac{\partial X^i}{\partial \sigma} \right)^2 \right\}  \label{eq:GGRTL}
\end{equation}
The condition of constant $P^+$-momentum per unit length means that the ``length" of the string is proportional to the total value of $p^+$.  The strings satisfy the boudnary condition that the slope, $\frac{\partial X^i}{\partial \sigma}$, is zero at the ends.
The GGRT choice of co-ordinates is not, of course, Lorentz covariant,  and the authors construct generators for the nontrivial Lorentz transformations.  The  transformations
$M_{i+}$, which change $X^+$,  are the difficult ones, since the condition $\tau=X^+$ has to be reestablished.   This is done by making a pseudo-conformal transformation of the $\sigma-\tau$ co-ordinates,  since the action is pseudo-conformally invariant.  The  Lorentz generators for the non-trivial transformations are trilinear.

Quantization of the model with the Lagrangian~(\ref{eq:GGRTL}) is straightforward.  The Hamiltonian is normal ordered,  and a constant term $\mu_0^2$ is added.   There is no problem with ghosts,  since the Lagrangian only involves transverse co-ordinates.  However,  Lorentz invariance is nontrivial.  The classical Lorentz generators do effect the required transformations,  but anomalous terms appear in the commutator between  transformations $M_{i+}$ and $M_{j+}$ when $i \neq j$.  These terms cancel when $d-2=24$ and $\mu_0^2=-1$,  which are thus the conditions for Lorentz invariance of the theory.

The Ramond and Neveu-Schwarz models were treated by Iwasaki and Kik-kawa~\cite{IK} in a similar way.  As might tbe expected,  the conditions for Lorentz invariance were $d-2=8$ and  $\mu_0^2=-\frac{1}{2}$ (Neveu-Schwarz) or $\mu_0^2=0$ (Ramond).

\section{Interacting Strings and Functional Integration}

\setcounter{equation}{0}

We now treat the subject of interacting strings by functional integration. 
The problem was initially considered by Gervais and Sakita~\cite{GS}.   The 
amplitudes constructed by them were not manifestly factorizable and the 
condition $d=26$ was not evident in their work.  The approach we shall 
describe~\cite{M1} was based on the light-cone GGRT string just treated.

It was realized independently by several people that a theory  of 
interacting strings could be constructed by allowing the free strings to split and join.  The Hilbert space now consists of any number 
of strings.  The free Lagrangian will be a sum of terms of the 
form~(\ref{eq:GGRTL}), one for each string,  together with an interaction which 
decreases or increases or decreases the number of strings by one.  The 
interaction vertex function will be zero unless the position  of the 
initial and final strings coincide.  In terms of operators, the vertex 
function $V$ will be the infinite product:
\begin{equation}
V=\prod_{\sigma , i} \delta \{ X_f^i (\sigma)  -  X_i^i (\sigma) \} 
\label{eq:vertex}
\end{equation}
Two strings can also meet at a point and recombine,  but we shall not 
encounter this term in our treatment.
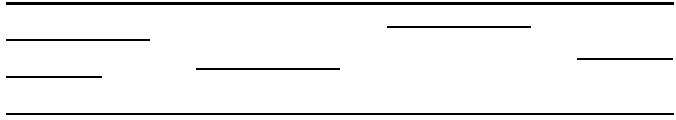
\begin{figure}
\begin{picture}(266,56)(-36, 0)

\put(7,7){\line(1,0){252}}

\put(7,21){\line(1,0){36}}

\put(7,35){\line(1,0){54}}

\put(7,49){\line(1,0){252}}

\put(79,24){\line(1,0){54}}

\put(151,40){\line(1,0){54}}

\put(223,28){\line(1,0){36}}

\end{picture}

\caption{Interacting string diagram}

\end{figure}

The string world sheet for a general interacting string process is shown 
in Figure 2.  The horizontal axis is the light-cone time  $\tau$,  the vertical axis is the length $\sigma$.  The $d-2$ transverse coordinates $X^i$ are 
orthogonal to the paper.  Thus three strings come in from $\tau =-\infty$, 
after a time the lower two join,  and so on until we reach $\tau = +\infty$. 
Note that the $X^i$ coordinates are discotntinuous actoss the horizontal 
lines,  since they are coordinates of different strings.  We have not 
separated the strings in the diagram to illustrate the fact that the total ``length" of all the strings, i. e., the total value of $P^+$, is constant.

For closed strings,  the ends of all strings in Figure 2 are identified,
so that the diagram consists of cylinders,  each cylinder representing a closed string propagating in time.  At the interaction points,  two closed strings join to form  one or one string splits into two.

The scattering amplitude  associated with the process shown will be found 
by functional integration of the transverse coordinates  $X^i (\sigma, 
\tau)$.  To begin,  we integrate separately over the parts of the world 
sheet associated with the different strings,  with a vertex function at 
each interaction time.  Since the vertex function requires  the $X^i$'s to 
be continuous across the interaction times,  we may simply integrate over the 
whole world sheet.  The $X^i$'s are discontinuous over the horinzontal 
lines of figure 2, and. at the horizontal lines, they will satisfy Neumann 
boundary conditions, namely that their derivates orthogonal to the 
boundary are zero.

Before performing the functional integration,  we continue analytically to
Euclidean time.  Thus Eq. (\ref{eq:GGRTL}) now becomes
\begin{equation}
\mathcal{L}=-\frac{1}{4\pi} \left\{ \left( \frac{\partial
X^i}{\partial \tau} \right)^2   +\left( \frac{\partial X^i}{\partial
\sigma} \right)^2 \right\}  \label{eq:L}
\end{equation}

After performing the functional integration,  we integrate over the 
intermediate times in figure 2, which we denote by $\tau_1,...,\tau_r$. 
The scattering amplitude for unexcited incoming incoming and outgoing 
particles is thus:
\begin{eqnarray}
A=\mathcal{N} \int \di \tau_1 ...[ \di \tau_i]... \di \tau_R \int \prod \di X^i(\sigma , 
\tau ) \hspace{2.in} & & \nonumber  \\   \times\exp \left\{i\sum_r 
\frac{1}{\pi \alpha_r}p_r^i \int \di \sigma X^i(\sigma, \tau_0) +\int \di 
\sigma \di \tau \mathcal{L} (\sigma ,\tau) -\sum_r p_r^- \tau_0 \right\} 
& &\label{eq:FI}
\end{eqnarray}
The factor $\mathcal{N}$ is a normalization factor,  which is a product of 
an overall normalization factor and a factor for each incoming or outgoing 
particle.  By considering the special case of a free string,  it can be 
shown to be unity.  The first sum in the exponent is over all incoming and 
outgoing particles; it is the wave function for the zero mode of the 
state.  The factor $\pi \alpha_r$ is the ``length" of the $r$th string, 
and the integral is over the string.  The time $\tau_0$ is the  time at 
the beginning or end of Figure 2 (large negative or large positive), 
depending on whether the particle is incoming or outgoing.  The wave 
functions of the nonzero modes of the string have been omitted,  since 
they will simply contribute to the normalization factor $\mathcal{N}$ if 
the particle is in its unexcited state.  They must be included for general 
excited particles.  The square brackets around $\tau_i$ indicate that one  $\tau$ is omitted from the variables of integration, as the entire amplitude is invariant under time translation.  The integration over this  variable simply gives us conservation of energy.  There are thus $N-3$ variables of integration. 

Since the functional integrand in (\ref{eq:FI}) is a quadratic function of 
the $X^i$, the functional integral can be performed by standard methods in 
terms of the reciprocal of the Laplacian and its determinant.  We define 
the Neumann function for the string world sheet by the equation:
\begin{equation}
\left\{ \left( \frac{\partial}{\partial \sigma} \right) ^2 + \left( \frac{ 
\partial}{ \partial \tau} \right) ^2 \right\} N (\sigma,\tau; \sigma^\prime,
 \tau^\prime)=2\pi \delta (\sigma -\sigma ^\prime)\delta (\tau  - 
\tau ^\prime ) -1/A \label{eq:Neumann}
\end{equation}
with Neumann booundary conditions,  i. e., the $\sigma$ deriatives of $N$ must be zero at the horizontal lines of Figure 2. $A$ is the area of the string diagram, with large finite initial and final times.
The last term is to take into account the zero mode,   since the inverse 
of the Laplacian operator can only be defined if the zero mode is 
excluded.\footnote{We are here improving on our original treatment of the 
extra terms on the right of (\ref{eq:Neumann}) . See~\cite[page 170] {Pol} 
.  As will  probably be evident in what follows,  the final results are 
unaffected.} Even though this term tends to zero as the initial and final 
times approach infinity it must be included,  since we really have to take the functional integral with finite initial and final times,  and then let the times approach $\pm \infty$.

On performing the functional integral,  we  obtain the result:
\begin{eqnarray}
  A=\int \di \tau_1 ...[\di \tau_i]... \di \tau_R  \Delta^{(d-2)/2} \hspace{2.4in}  & & \nonumber \\ \times \exp
\left\{\sum_{r,s} \frac{1}{\pi^2 \alpha_r \alpha_s}p_r^i p_s^i \int \di 
\sigma \di \sigma^\prime N(\sigma, \tau_0 ; \sigma^\prime 
,\tau_0^\prime)-\sum_r p_r^- \tau_0 \right\}
\label{eq:FI2}
\end{eqnarray}
The factor $\Delta$ is $-1/2\pi$ times the determinant of the Laplacian on 
the string diagram.  The first summation within the braces is now over all pairs of strings, and the $\sigma$-integtal is over the string at $\tau=\pm\infty$.
If the two strings are different (and at $\pm \infty$),  the Neumann function will be independent of the position along the string, so that the $\sigma$ integrals may be written $N(r,s)$, 
where r and s denote the positions of the $r$th and $s$th string.  The 
integrals where $r$ and $s$ refer to the same string contribute 
normalization factors,  independent of the shape of the string diagram. 
We denote the product of all these normalization factors by $N$.
Furthemore, it is not difficult to show that the last term in the braces 
of equation (\ref{eq:FI2}) simply changes the $(d-2)$-vector product $p_r^i 
p_s^i$ into a d-vector product.  On inserting these changes in equation (\ref{eq:FI2}), we obtain our final result:
\begin{equation}
A=\int \di \tau_1 ...[\di \tau_i]... \di \tau_R 
\Delta^{(d-2)/2}N\exp \left\{-2\sum_{r>s} p_r.p_s N(r,s) \right\}.
\label{eq:FI3}
\end{equation}

As we have been using the light-cone frame our approach,  while manifestly unitary in a positie-definite Hilbert space,  is not manifestly Lorentz invariant. (While the momentum-dependent factor in equation (\ref{eq:FI3}) is Lorentz covariant, the ``lengths" of the strings in Figure 2 are proportional to $P^+$ and therefore dependent on the Lorentz frame.)  As we have already mentioned, GGRT constructed Lorentz generators for the free string and showed that they had the correct properties if $d=26$.  By applying these generators to the vertex (\ref{eq:vertex}),  we have shown that the interacting-string theory is Lorentz invariant if $d=26$ \cite{M1} \cite{M2}.

Now let us see how we can obtain the result quoted in the introduction for the Born term of the string model.  In that case the only horizontal lines in Figure 2 will be those from the external states,  i. e., the string world sheet will have genus zero.  Since the Neumann functions are conformally invariant, they can be evaluated by conformally transforming  the string diagram onto the upper half plane, with the external particles transforming onto points on the real axis.  The formula for doing so is a special case of the Schwarz-Christoffel transformation, namely
\begin{equation}
\rho = \sum_{r=1}^{n} \pi\alpha_r \ln(z-Z_r) \label{eq:CT}
\end{equation}
The variable $\rho$ $(=\tau + i\sigma)$ is the co-ordinate on the string diagram, $z$ the co-ordinate on the upper half-plane. The string lengths $\pi\alpha$ are considered positive for incoming strings, negative for outgoing strings.  The $r$th string at $\tau =\pm\infty$ transforms to the point $Z_r$ on the real axis.  It is often  convenient to take $Z_N=\infty$.

The Neumann function for the upper half-plane is simply
\begin{equation}
N(z,z^\prime)=\ln|z-z^\prime| + \ln|z-z^{\prime \ast}|  \label{eq:Neu}
\end{equation}
(For closed strings,  we transform the string diagram onto the whole plane. The strings at $\tau=\pm \infty$ transform to points anywhere in the plane, and the variables $Z_r$ are complex.  The second term on the right of (\ref{eq:Neu}) is absent.)

The momentum-independent factors of (\ref{eq:FI3}), i. e., the factors to the left of the exponential, may be treated in one of two ways.  The simplest method is to make use of the proved Lorentz inariance of the amplitude in the critical dimension.  We take the variables $P^+$ for all but two of the strings,  one outgoing and one incoming, each at the bottom of Figure 2, to be equal to zero.  Since the lengths of the strings are proportional to $P^+$, this means that the string diagram consists of one long string with several infinitely short stings entering or exiting at the top.  The vertex functions, and therefore the momentum-independent factor,  are then easily shown to be unity.  Thus, on transforming the integral (\ref{eq:FI3}) onto the upper half-plane, inserting the formula (\ref{eq:Neu}) for the Neumann functions, and replacing the $N-3$ variables of integration $\tau_r$ by the corresponding real $Z_r$'s, we easily obtain the formula (\ref{eq:KN}). (For closed strings the $Z_r$'s are integfated over the entire comple plane).

The foregoing analysis is very simple,  but it canot be extended to loops.  An alternative method is to treat the string diagram in Figure 2 directly,  without using Lorentz invariance \cite[p. 66]{UST}.  We then have to examine the factor $\Delta$ in (\ref{eq:FI3}),  the determinant of the Laplacian.  For the upper half-plane,  this factor is of course a constant;  however, we require it for the string diagram and, since it requires regularization,  it is \emph{not} invariant under a conformal transformation.  The  change of $\Delta$  under such a transformation (known to physicists as the ``conformal anomaly") has been worked out by McKean and Singer \cite{MS} (see also Alvarez \cite{A}). On applying their result to the conformal transformation from the string diagram to the upper half-plane,  and calculating the Jacobian from the $\tau_r$'s to the $Z_r$'s,  which is not as simple as it was in the first method,  we obtain equation(\ref{eq:KN}), if and only if $d-2=24$.

The calculation of the one-loop amplitude by functional integration involves
new considerations, since the string-diagram world sheet  is  then conformally equivalent to an annulus instead of the upper half-plane.  (The closed-string world sheet is conformally equivalent to a torus).  For details of the calculation, we refer the reader to refs. \cite{PR}, \cite{Arf} \cite[p.72]{UST}.  

Multiple-loop string diagrams for processes involving open and closed strings are conformally equivalent to  Riemann suffaces of higher genus,  possibly with holes if some of the strings are open. Such  amplitudes  are most easily treated by considering their analytic properties in the moduli space of Riemann surfaces of genus $g$.  The analytic properties have been examined by Belavin and Knizhnik \cite{BK}.  who are mainly interested in the Polyakov formalism,  to be outlined in the final section of this article, but their methods can also be used in the present context.  More precisely, we take a conformal metric which is the product of an analytic function on the Riemann surface and its conjugate complex,  as has been suggested by Sonoda \cite{So}.  Such a metric necessarily has $2(g-1)$ zeros, which are of course singular points.  It can then be showm that $\Delta$, evaluated in this metric, is an analytic function in moduli space except for known singularities where the moduli space degenerates or where two zeroes of the metric coincide.

We should point out that the whole of this section is,  strictly speaking,
incorrect, since we have made an illegal Wick rotation in continuing to imaginary time.  A Wick rotation cannot be made if there is an intermediate state with lower energy than the initial and final states, since the integral over the exponential $\exp \{-i(E_{init} - E_{int})t\}$ diverges.  Normally such an intermediate state is subtracted out and treated explicitly,  but such a procedure causes difficulties if applied to string models.  In fact,  all the amplitudes we have written here are real, and they diverge if the external energy in any channel is sufficiently large.  We have already mentioned this fact in the introduction;  we now see the reason for it from the point of view of string models.  In the early work on dual models,  the problem was treated by analytic continuation from sufficiently low energies.
For closed-string amplitudes,  it is not always possible to find a region where the energies in all channels are sufficiently low.  Nevertheless, D'Hoker and Phong \cite{DP} have shown that the integral for the one-loop closed-string amplitude can be divided into several parts,  each of which can be analytically continued.  They thereby obtain a finite amplitude with the correct singularities. It would be preferable to perform the functional integration in such a way that the amplitude is finite and unitary as it stands,  without the necessity of analytic continuation.  This can be done by undoing the illegal Wick rotation, as has been shown by Berera \cite{Be}, again for the one-loop closed-string amplitude.  It is not necessary to undo
the entire Wick rotation; one need only consider the regions where the moduli space degenerates.  One then obtains an amplitude which is finite and 
unitary.

\section{Functional Integration for the RNS Model}

\setcounter{equation}{0}

In this section we shall summarize briefly how the functional integration of the previous section can be applied to the Ramond-Neveu-Schwarz model \cite{RNS}.  The advantage of the present trestment is that the Ramond (R) and Neveu-Schwarz (NS) models now appear much more directly as two sectors of a single model.

In addition to the commuting variables of the previous model,  we  have two anticommuting variables, $S_1^i$ and $S_2^i$ with (transverse) vector indices, corresponding to the anticommuting operators in the operator treatment of the RNS model. The Lagrangian will be the sum of two terms, one being the same as the Lagrangian (\ref{eq:L}) of the previous model, and the other being given by the formula
\begin{equation}
\mathcal{L}_2=-\frac{1}{2\pi}\left\{S_1\left(\frac{\partial}{\partial\tau}+i\frac{\partial}{\partial\sigma}\right)S_1+S_2\left(\frac{\partial}{\partial\tau}-i\frac{\partial}{\partial\sigma}\right)S_2\right\}.
\label{eq:L2}
\end{equation}

The boundary conditions at the end of the string are either $S_1 = +S_2$ 
or $S_1 =- S_2$.  For incoming fermions or outgoing anti-fermions (R) we have 
the plus sign at both ends; for incoming anti-fermions or outgoing 
fermions  we have the minus sign at both ends,  while for bosons (NS) we have 
different signs at the two ends.  The boundary conditions are continuous 
along the top or bottom of the horizontal lines in Figure 2, but one sign
changes into the other at each joining point.  When expanding the $S$'s in 
normal modes $b^i$,  we use the interval $-\pi\alpha \leq \sigma \leq 
\pi\alpha$,  where $\pi\alpha$ is, as usual, the length of the string,  and we 
take $S_2^i(-\sigma) =S_1^i(\sigma), \sigma>0$.  Thus the bosons have
half-integral mode numbers, the fermions integral mode numbers.  With closed 
strings $S_1 $and $S_2$ change sign after one complete rotation for bosons
and do not change sign for fermions.  Thus the closed-string model has four sectors,
NS-NS, NS-R, R-NS and R-R.

The vertex function is no longer given simply by the overlap integral,  but there is an extra factor $G$ given by the equation:
\begin{equation}
G=-\lim_{\sigma \rightarrow \sigma_1}i^{\pm 1/4}(\sigma - \sigma_1)^{3/4}S_1^i (\sigma)\left (\frac{\partial}{\partial\tau} + i\frac{\partial}{\partial\sigma} \right )X^i(\sigma)
\label{eq:G}
\end{equation}
The sign in the factor $i^{\pm 1/4}$ is plus or minus depending on wether the strings join or separate.  The expression is evaluated at a point $\sigma$ near the interaction point $\sigma_1$.
A factor $G$ at each vertex will of course appear  in the functional integrand. The factor $G$ is necessary in order to prove Lorentz invariance, and in order that the functional integration give the same result as the operator formalism.

As in the Bose case,  the functional integration can be performed in terms of Neumann functions.
For tree-level amplitude, the result can be conformally transformed onto the upper half-plane.
Here we must bear in mind that the operators $S^i$, and therefore the Neumann functions involving them, have conformal weight 1/2, i. e., on transforming to the $z$-plane we must include an extra factor $(\partial\rho/\partial z)^{-1/2}$ for $S_1$ and the complex conjugate of this factor for $S_2$  We omit the details of the calculation.  The result for scattering amplitudes of bosons is the same as that given by the operator calculation in the NS model.  We can also calculate the scattering amplitude for fermion-antifermion scattering.  This was the first such result  as, in the operator formalism,   amplitudes involving interacting Ramond fermions were much more difficult to treat than those involving bosons.  The rather complicated
calculations in the operator fromalism  were completed soon after by Schwarz and Wu \cite{SW},
following work by Thorn, Corrigan and Olive; Olive and Scherk; and Brink, Olive, Rebbi and Scherk \cite{FEV}.  Their results agreed with those calculated by functional integration.

Berkovits \cite{Be1} showed that the treatment of the RNS model outlined in this section was equivalent to a functional integration over a super-worldsheet.  Superfields had previously been applied to the RNS model by Fairlie and Martin \cite{FM} and Brink and Winnberg \cite{BW},  and also by Polyakov\cite{P1} in the fromalism to be mentioned in the next section.  The use of supersheets
introduces a considerable simplification,  since there is no longer an operator $G$ at the interaction points.

\section{Comparison with the Polyakov Formulation for Functional Integration}

\setcounter{equation}{0}

For completeness we shall now compare briefly the formulation of the last two
sections  with another formulation proposed later by Polyakov \cite{P1}, even
though it takes us beyond the time frame of this volume. (For a very much more detailed treatment of the Polyakov model, see ref. \cite{Pol}.) Instead of starting
from the Nambu-Goto string, Polyakov starts from an action defined by Brink,
DiVecchia and Howe \cite{BDH} and by Deser and Zumino \cite{DZ}.  One
introduces a general co-ordinate system on the world sheet,  with
co-ordinates  $(\sigma_a, \sigma_b $) and  metric $g_{ab}$;
the position of the point $\sigma$ in $d$-dimensional space is denoted by
$X^\mu  (\sigma)$. The Lagrangian is then
\begin{equation}
\mathcal{L}=-\frac{1}{4\pi}g^{1/2 }g_{ab} \left( \frac{\partial
X^\mu}{\partial \sigma_a} \frac{\partial X_\mu}{\partial \sigma_b } \right).
  \label{eq:LP}
\end{equation}

To calculate the $S$-matrix with $N$ external ground-state closed strings in
$g$th order perturbation perturbation theory, Polyakov considers a Riemann
surface of genus $g$ with $N$ punctures (which, as we have seen, is
conformally equivalent to a surface with the strings going to $\pm \infty$).
At each puncture there is a factor $\exp \{ ip_r^{\mu} X_{\mu} (\sigma)\}$.
The functional integrand is thus
\begin{equation}
\prod_{r=1}^{N}\int \di^2 \sigma_r g^{1/2}(\sigma_r) \exp\left\{i\sum_{r}
p_r^\mu X_\mu(\sigma_r)+\int \di^2\sigma\mathcal{L}(\sigma)\right \}.
\label{eq:FInt}
\end{equation}

Polyakov now functionally integrates over the variables $X^\mu(\sigma)$ and
$g_{ab}(\sigma)$. The integration over $X^\mu (\sigma)$ proceeds as before. To 
integrate over $g_{ab}$, we first notice that we can use an Einstein
co-ordinate transformation 
to convert the metric to a covariantly constant metric $g\delta_{ab}$
(in Euclidean space).  Due to invariance under such transformations,
this integration can be reduced to a functional Faddeev-Popov determinant
which will contain a conformal anomaly similar to the conformal anomaly 
in the determinant of the ordinary Laplacian mentioned in section 4.  Unless
$d=26$,  the conformal annomaly will appear as an extra field when
integrating over the remaining $g$,  and we obtain a string theory 
together with an extra Liouville field.  This noncritical string theory
has been studied to a certain extent, but not nearly as much as that
with the critical dimension $d=26$.  For this dimension the conformal
anomalies in the two functional determinants cancel,  and the theory 
possesses a Weyl conformal invariance,  i. e., an invariance under a 
conformal change of $g$ without a compensating change in the co-ordinate
system.  ``Most" of the integration over the remaining variable $g$ will 
therefore simply be as integration over gauges and,  in fact, no further
Faddeev-Popov determinant is introduced.  However, if the genus
is not zero,  not all Riemann surfaces of the same genus are conformally 
equivalent,  and we are left with an integration over the variables
charecterizing the conformal class.

Not surprisingly, the Polyakov integral for genus zero reproduces the formula (\ref{eq:KN}). Again, the amplitude is projectively invariant,  so that the positions of three external particles must be fixed at arbitrary values.

The Polyakov formulation has the advantage over the light-cone forulation of section 4 of being relativistically covariant. Also, the integragion variables characterizing the conformal classes of Riemann surfaces correspond to the parametrization of such classes by Bers \cite{Bers}.  Thus the Polyakov formulation can be related to the mathematical theory of Riemann surfaces.
As the conformal anomaly cancels,  one can avoid reference to the metric (though it may not always be advantageous to do so \cite{DS} \cite{So}).
On the other hand, the Polyakov formulation gives us only the $S$-matrix; one cannot go off shell, whereas the light-cone formulation gives us a complete quantum mechanical theory where we are not restricted to infinite times and can go off shell,  allbeit noncovariantly.  Also, the Polyakov formulation treats different orders in perturbation theory separately and unitarity,  even perturbative unitarity,  is far from obvious. In fact two inequivalent theories with the same Born term cannot both be unitary.  It is therefore important  that D'Hoker and Giddings \cite{DG} have shown that the two formulations are in fact equivalent;  they are different gauges of the same theory.  D'Hoker and Giddings point out that it should be possible to prove unitarity directly in the Polyakov formulation (for a possible approach see 
\cite[Chapter 9]{Pol}),but such a proof appears to be considerably more complicated that
in the light-cone formulation,  where unitarity is manifest.

Thus, with the two gauges of the functional integration theory, we can more easily see different aspects of the system than we could with each gauge separately.
 
Polyakov also considered the RNS string, and most of the general remarks we have made apply to this case as well.  Aoki, D'Hoker and Phong \cite{ADP} have treated the unitarity of the Polyakov RNS string by relating it to Berkovits'
light-cone formulation with a super-worldsheet, mentioned in the previous section, which is manifestly unitary, but some problems remain with external fermions \cite{G2}.
 
The RNS formalism has space-time supersymmetry after projecting out half of
the states using the Gliozzi-Scherk-Olive (GSO) projection \cite{GSO} (which 
we have not distussed in this article).  However,  it is not \emph{manifestly}
supersymmetric.  Green and Schwarz \cite{GS2},  using the light-cone gauge,
have given a formulation with manifest space-time supersymmetry.  This makes
it much easier to show that the principal divergence of the theory,  namely the 
``dilaton" divergence,  does not occur \cite{MF},  as the absence of this 
divergence depends on space-time supersymmetry.  The Green-Schwarz formalism 
requires operators at the interaction points of Figure 2, and calculations, 
especially of multi-loop amplitudes,  are thereby made more difficult..  At the
cost of some complication, Berkovits \cite{B2} has given a manifestly 
Lorentz-covariant formulation of the Green-Schwarz theory,  which includes 
the ``ghosts" associated with the Faddeev-Popov determinant discussed above.  His
formulation thus possess manifest Lorentz covariance and manifest space-time 
supersymmatry, and avoids the necessity of operators at the interacting points. 
As before, unitarity is established by comparing the theory with the light-cone
formulation of the previous section.


\begin{thebibliography}{99}
\bibitem{Ven}  G. Veneziano, Nuovo Cimento 57A (1968) 190.
\bibitem{Vir}  M. A. Virasoro,  Phys. Rev. 177 (1969) 2309.
\bibitem{Sh}   J. A. Shapiro, Phys. letters 33B (1970) 361.
\bibitem{BRV} K Bardak\c{c}i and H. Ruegg, Phys Letters 28B (1968) 342\\ M.A. Virasoro, Phys. Rev. Letters 22 (1969).
\bibitem{n-point}  K. Bardak\c{c}i and H. Ruegg, Phys. Rev 181 (1969) 1884\\ H. M. Chan and S. T. Tsou, Phys. Letters 28B (1969) 485\\ C. G. Goebel and B. Sakita,  Phys. Rev. Letters 22 (1969) 257 \\Z. Koba and H. B. Nielsen Nucl. Phys B10 (1969) 63.
\bibitem{KN} Z. Koba and H. B. Nielsen, Nucl. Phys B12 (1969) 517.
\bibitem{BMFV} K. Bardak\c{c}i and S. Mandelstam, Phys.Rev 184 (1969) 1640\\S. Fubini and G. Veneziano Nuovo Cimento 64A.
\bibitem{FGV} S.Fubini, D. Gordon and G. Veneziano, Phys. Letters 29B (1969). 679\\Y. Nambu, Proc. Intern. Conf. on Symmetries and Quark Models, Wayne State University (1969)\\L. Susskind, Phys. Rev. D1 (1970) 1182.
\bibitem{R} P. Ramond, Phys. Rsv D3 (1971) 2415.
\bibitem{NS} A. Neveu and J. H. Schwarz, Nucl. Phys. B31 (1971) 86 \\A. Neveu, J. H. Schwarz and C. B. Thorn, Phys. Letters 35B (1971 529.
\bibitem{KSV} K. Kikkawa, B.Sakita and M. A. Virasoro, Phys. Rev. 184 (1969), 1701
\bibitem{one-loop} D. Amati, C. Bouchiat and J-L. Gervais, Nuovo Cimento Letters 2 (1969) 399\\ K. Bardak\c{c}i, M. B. Halpern and J. A. Shapiro, Phys. Rev.185 (1969) 1910\\K. Kikkawa, B. Sakita, G. Veneziano and M. A. Virasoro,  footnote to ref.~\cite{KSV}.
\bibitem{KT} M. Kaku and C. B. Thorn, Phys. Rev. D1 (1979) 2860.
\bibitem{GNSS} D. J. Gross, A. Neveu, J. Scherk and J. H. Schwarz, Phys. Rev D2 (1970), 697.
\bibitem{n-loop} V. Alessandrini, Nuovo Cimento 2A (1971) 321 \\M. Kaku and L. P. Yu, Phys. Letters 33B (1970) 166; Phys. Rev. D3 (1971) 2992, 3007, 3020 \\ C. Lovelace, Phys. Letters 32B (1970) 703.
\bibitem{AA} V. Alessandrini and D. Amati, Nuovo Cimento 4A (1971) 793.
\bibitem{FN} D. B. Fairlie and H. B. Nielsen, Nucl. Phys. B20 (1969) 637.
\bibitem{FS} G. Frye and L. Susskind, Phys. Letters B31 (1970) 537.
\bibitem{L2} C. Lovelace, Phys Letters B34 (1971) 500.
\bibitem{Vir2} M. A. Virasoro, Phys. Rev D1 (1970) 2933.
\bibitem{DDF} E. Del Giudice, P. DiVecchia and S. Fubini, Ann. of Phys. 70 (1972) 378.
\bibitem{Bro} R. C. Brower, Phys. Rev. D6 (1972) 1655.
\bibitem{GT} P. Goddard and C. B. Thorn, Phys Letters 40B (1972) 235.
\bibitem{NSghost} R. C. Brower and K. A. Friedman, Phys. Rev. D7 (1973) 535\\J. H. Schwarz, Nucl. Phys. B46 (1972) 61.
\bibitem{string} Y. Nambu, Proc. Intern. Conf. on Symmetries and Quark Models, Wayne State University (1969)\\H. B. Nielsen, Proc. 15th Intern. Conf on High-Energy Physics, Kiev (1970)\\L. Sussind, Nuovo Cimento 69A (1970) 457.
\bibitem{Go} T. Goto, Prog. Theor. Phys. 46 (1971) 1560.
\bibitem{GGRT} P. Goddard, J. Goldstone. C. Rebbi and C. B. Thorn, Nucl. Phys. B56 (1973), 109.
\bibitem{IK} Y. Iwasaki and K. Kikkawa, Phys. Rev. D8, (1973), 440.
\bibitem{GS} J-L Gervais and B. Sakita, Phys. Rev Letters 30 (1973) 716.
\bibitem{M1} S.Mandelstam, Nucl. Phys. B64 (1973) 205.
\bibitem{Pol} J. Polchinski, String Theory,  Volume 1. Cambridge, 1998.
\bibitem{M2} S. Mandelstam, Nucl. Phys B83 (1974) 413.
\bibitem{UST} M. B. Green and D. J. Gross (Eds.), Unified String Theories (Proceedings of Workshop) World Scientific, 1986.
\bibitem{MS} H. McKean and I. M. Singer, J. Diff. Geom. 1 (1973) 332.
\bibitem{A} O. Alvarez, Nucl. Phys. B216 (1983) 125.
\bibitem{PR} S. Mandelstam, Phys.Reports 13 (1974), 260.
\bibitem{Arf} H. Arfaei, Nucl. Phys. 85, (1975), 535.
\bibitem{BK} A. A. Belavin and V. G. Knizhnik, Phys. Letters B168 (1986), 201.
\bibitem{So} H. Sonoda, Nucl. Phys B284 (1987), 157.
\bibitem{DP} E. D'Hoker and D. H. Phong, Phys. Rev. Letters 70 (1993), 3692.
\bibitem{Be} A. Berera, Nucl. Phys. B411 (1994) 157.
\bibitem{RNS} S. Mandelstam, Phys. Letters 46B (1973),447; Nucl. Phys. B69 (1964), 77.
\bibitem{SW} J. H. Schwarz and C. C. Wu, Phys. Letters 47B (1973), 453.
\bibitem{FEV}  C. B. Thorn, Phys Rev D4 (1971) 1112\\ E. F. Corrigan and D. Olive, Nuovo Cim. 11A (1972) 749\\ D. Olive and J. Scherk, Nucl. Phys B64 (1973), 334\\ L. Brink, D. Olive, C. Rebbi and J. Scherk, Phys. Lett. 45B (1943) 379.
\bibitem{Be1} N. Berkovits, Nucl. Phys. B276 (1986) 650 ; B304 (1988) 537.
\bibitem{FM} D. B. Fairlie and D. Martin, Nuovo Cim. 21A (1974) 647.
\bibitem{BW} L. Brink and J. O. Winnberg, Nucl. Phys. B103 (1976) 445.
\bibitem{P1} A. M. Polyakov, Phys. Letters (1981) 207; (1981) 211.
\bibitem{BDH} L. Brink, P. DiVecchia  and P.Howe, Phys. Letters 65B (1976),
471.
\bibitem{DZ} S. Deser and B. Zumino, Phys. Letters 65B (1976), 369.
\bibitem{Bers} L. Bers, Bull. Amer. Math. Soc. 5 (1981), 131.
\bibitem{DS} M. J. Dugan and H. Sonoda, Nucl. Phys B289 (1987) 227.
\bibitem{DG} E. D'Hoker and S. B. Giddings, Nucl. Phys. B291 (1987), 90.
\bibitem{ADP} K. Aoki, E. D'Hoker and D. H. Phong, Nucl. Phys B342 (1990) 149
\bibitem{G2} S. B. Giddings,  Comm. Math. Phys. 143 (1992) 355.
\bibitem{GSO} F. Gliozzi, J. Scherk and D. Olive, Nucl. Phys. B122 (1977), 253.
\bibitem{GS2} M. B. Green and J. H. Schwarz, Nucl Phys. B181 (1981), 502; B198 (1982) 252; B243 (1984) 285.
\bibitem {MF} S. Mandelstam, Phys. Letters  277B (1992) 86.
\bibitem{B2} N. Berkovits Phys. Letters 232B (1989) 184; 300B (1993) 53; Nucl.
Phys. B358 (1991) 169; B395 (1993) 77.



\end{thebibliography}
\end{document}